# All-optical complex field imaging using diffractive processors

Jingxi Li[1,2,3], Yuhang Li[1,2,3], Tianyi Gan[1,3], Che-Yung Shen[1,2,3], Mona Jarrahi[1,3], and Aydogan Ozcan[1,2,3*]

[1]Electrical and Computer Engineering Department, University of California, Los Angeles, CA, 90095, USA

[2]Bioengineering Department, University of California, Los Angeles, CA, 90095, USA

[3]California NanoSystems Institute (CNSI), University of California, Los Angeles, CA, 90095, USA

[*]Correspondence to: ozcan@ucla.edu

## Abstract

Complex field imaging, which captures both the amplitude and phase information of input optical fields or objects, can offer rich structural insights into samples, such as their absorption and refractive index distributions. However, conventional image sensors are intensity-based and inherently lack the capability to directly measure the phase distribution of a field. This limitation can be overcome using interferometric or holographic methods, often supplemented by iterative phase retrieval algorithms, leading to a considerable increase in hardware complexity and computational demand. Here, we present a complex field imager design that enables snapshot imaging of both the amplitude and quantitative phase information of input fields using an intensity-based sensor array without any digital processing. Our design utilizes successive deep learning-optimized diffractive surfaces that are structured to collectively modulate the input complex field, forming two independent imaging channels that perform amplitude-to-amplitude and phase-to-intensity transformations between the input and output planes within a compact optical design, axially spanning ~100 wavelengths. The intensity distributions of the output fields at these two channels on the sensor plane directly correspond to the amplitude and quantitative phase profiles of the input complex field, eliminating the need for any digital image reconstruction algorithms. We experimentally validated the efficacy of our complex field diffractive imager designs through 3D-printed prototypes operating at the terahertz spectrum, with the output amplitude and phase channel images closely aligning with our numerical simulations. We envision that this complex field imager will have various applications in security, biomedical imaging, sensing and material science, among others.

# Introduction

Optical imaging can characterize diverse properties of light, including amplitude, phase, wavelength, and polarization, which provides abundant information about samples, such as their morphology and composition. However, conventional image sensors and focal plane arrays, based on e.g., Complementary Metal-Oxide-Semiconductor (CMOS) or Charge-Coupled Device (CCD) technologies, are inherently constrained to detecting only the intensity of the optical field impinging on their active area. Measuring the phase information of a complex field presents challenges, which require indirect encoding through interferometric or holographic detection systems[1–3]. Some of the traditional examples of phase imaging techniques include Zernike phase contrast microscopy and interferometric microscopy. Subsequent developments of quantitative phase imaging (QPI) have enabled high-precision characterization of phase information; advancements such as the Fourier phase microscopy[4], Hilbert phase microscopy[5] and digital holographic microscopy[6–12] have also emerged, making QPI a potent label-free optical measurement technique. Nevertheless, these QPI methods often necessitate relatively bulky experimental setups and rely on iterative algorithms based on multiple measurements to digitally reconstruct the desired phase information, leading to slow imaging speeds.

Recently, fueled by the advances made in deep learning, the application of deep neural networks has been adopted for accurate and rapid reconstruction of phase information in complex fields through a single feed-forward operation[13–20]. While these deep learning-based approaches offer considerable benefits, they typically demand intensive computational resources for network inference, requiring the use of graphics processing units (GPUs). Simultaneously, the progress in micro- and nano-fabrication technologies facilitated the development of metasurfaces[21–29] and thin-film optical components[30,31] for QPI applications. However, the functionality of these devices still relies on indirect encoding processes that generate intensity variations on the sensor plane, such as diffused speckles[21] or polarized interference patterns[29]. These existing solutions, therefore, necessitate digital computation for image reconstruction and, in certain cases, also require the incorporation of additional hardware along the optical path, such as polarizers and polarization cameras.

In this work, we demonstrate the design of a complex field imager that can *directly* capture the amplitude and phase distributions of an incoming field using an intensity-only image sensor array. As shown in **Fig. 1a**, this complex field imager is composed of a series of spatially engineered diffractive surfaces (layers)[32–49], optimized using supervised deep learning algorithms to simultaneously perform two tasks: (1) an amplitude-to-amplitude (A→A) transformation and (2) a phase-to-intensity (P→I) transformation. Here, the first task involves mapping the amplitude of the incoming complex field to a specific output field of view (FOV) that is solely dedicated to amplitude imaging, independent of the input wave's phase profile. The second task, on the other hand, aims to approximate a nonlinear transformation by converting the phase of the incoming wave into an intensity pattern at another output FOV, exclusively used for quantitative phase imaging, QPI. Therefore, by placing complex objects or feeding complex fields into the input FOV of the diffractive complex field imager and measuring the intensity distributions at its output FOVs, the amplitude and phase information of the input complex objects/fields can be *directly* obtained within a *single* intensity-only image recording step, eliminating the necessity for any form of image reconstruction algorithms. In addition to this spatially multiplexed design of the diffractive complex field imager, termed design I (**Fig. 1a**), we also explored additional complex field imager designs by incorporating wavelength multiplexing. As illustrated in **Fig. 1b** and **c**, these wavelength multiplexed designs, termed designs II and III, operate by detecting the output amplitude and phase signals at two distinct wavelengths ($\lambda_1$ and $\lambda_2$, respectively). The difference between designs II and III is that the design II utilizes a shared FOV for both the amplitude and phase channel outputs, while the design III maintains two spatially separated FOVs, each dedicated to either the output amplitude or phase images.

After completing the training of these different designs, we blindly tested the performance and generalization capabilities of our trained diffractive complex field imager models. We quantified their imaging errors using thousands of examples of input complex fields, each composed of independent information channels encoded in the amplitude and phase of the input field. The results demonstrated that our diffractive models could successfully generalize to new, unseen complex test fields, including those with structural features distinctly different from the training objects. Through numerical simulations, we further analyzed the spatial resolution and sensitivity of both the amplitude and phase channels of our diffractive complex field imagers. These analyses revealed that our designs could resolve amplitude features with a linewidth of $\geq 1.5\lambda_m$ and phase features with a linewidth of $\geq 3\lambda_m$, where $\lambda_m$ represents the mean wavelength. Furthermore, our studies showed that by integrating an additional diffraction efficiency-related loss term into the training function, one could achieve diffractive imager models with enhanced output power efficiencies with minimal compromise in imaging performance.

Apart from these numerical analyses, we also conducted an experimental proof-of-concept demonstration of our diffractive complex field imagers using the terahertz part of the spectrum by fabricating the resulting diffractive layers using 3D printing. For our experiments, we constructed test objects (never seen during the training) with spatially structured amplitude or phase distributions through 3D printing and surface coating techniques. Our experimental results successfully reconstructed the amplitude and phase images of the test objects, closely matching our numerical simulations and the ground truth, validating the effectiveness of our diffractive complex field imager designs. While our experimental demonstrations were conducted in the terahertz spectrum, our designs are scalable and can be adapted to other spectral bands by scaling their dimensions proportional to the wavelength of operation. The compact size of our diffractive designs, with an axial span of ~$100 \times \lambda_m$, facilitates easy integration into existing optical imaging systems and focal plane arrays that operate at different parts of the electromagnetic spectrum. This complex field imager design also does not include any components that are sensitive to the polarization of light, maintaining its amplitude and phase imaging function regardless of the input polarization distribution of the input field. Given all these advantages, including the small footprint, speed of all-optical computation and low-power operation, we believe that this all-optical complex field imaging approach will find broad applications in e.g., defense/security, biomedical imaging, sensing and material science.

## Results

**Designs of diffractive complex field imagers**

**Figure 1a** illustrates a spatially multiplexed design of our diffractive complex field imager, termed the design I. This diffractive imager is composed of 5 diffractive layers (i.e., $L_1$, $L_2$, ..., $L_5$), where each of these layers is spatially coded with 200×200 diffractive features, with a lateral dimension of approximately half of the illumination wavelength, i.e., ~$\lambda/2$. These diffractive layers are positioned in a cascaded manner along the optical axis, resulting in a total axial length of $150\lambda$ for the entire design. A complex input object, $i(x,y) = A(x,y)e^{j\phi(x,y)}$, illuminated at $\lambda$ is placed at the input plane in front of the diffractive layers. This complex object field exhibits an amplitude distribution $A(x,y)$ that has a value range of [$A_{DC}$, 1], along with a phase distribution $\phi(x,y)$ ranging within [0, $\alpha\pi$]. Here, $A_{DC}$ denotes the minimum amplitude value of the input complex field, and $\alpha$ is the phase contrast parameter of the input complex field. Without loss of generality, we selected default values of $A_{DC}$ and $\alpha$ as 0.2 and 1, respectively, for our numerical demonstrations. Note that it's essential to work with $A_{DC} \neq 0$ since otherwise the phase would become undefined. After the input complex fields are collectively modulated by these diffractive layers $L_1$-$L_5$, the resulting optical fields $o(\lambda)$ at the output plane are measured by the detectors within two spatially separated output FOVs, i.e., $FOV_{Phase}$ and $FOV_{Amp}$, which produce intensity distributions $|o_{Phase}(\lambda)|^2$ and $|o_{Amp}(\lambda)|^2$ that correspond to the phase and amplitude patterns of each input complex field, respectively. In addition, we also defined a reference signal region $\mathcal{R}$ at the periphery of the $FOV_{Phase}$, wherein the average measured intensity across $\mathcal{R}$ is used as the reference signal $R(\lambda)$ for normalizing the quantitative phase signal $|o_{Phase}(\lambda)|^2$. This normalization process is essential to ensure that the detected phase information is independent of the input light intensity fluctuations, yielding a quantitative phase image $O_{Phase}(\lambda) = \frac{|o_{Phase}(\lambda)|^2}{R(\lambda)}$, regardless of the diffracted output power. Overall, the objective of our training process is to have the phase image channel output approximate the ground truth phase distribution of the input complex field, i.e., $O_{Phase}(\lambda) \approx \phi(\lambda)$, demonstrating an effective phase-to-intensity (P→I) transformation. Concurrently, the training of the diffractive layers also aims to have the diffractive output image in the amplitude channel, i.e., $|o_{Amp}(\lambda)|$, proportionally match the ground truth amplitude distribution of the input complex field after subtracting the amplitude DC component $A_{DC}$, i.e., $|o_{Amp}(\lambda)| \propto (A(\lambda) - A_{DC})$, thereby achieving a successful amplitude-to-amplitude (A→A) transformation performed by the diffractive processor.

In addition to the spatially multiplexed design I described above, we also created an alternative complex field imager design named design II by incorporating *wavelength multiplexing* to construct the amplitude and phase imaging channels. As illustrated in **Fig. 1b**, this approach utilizes a dual-color scheme, where the amplitude and phase of the input images are captured separately at two distinct wavelengths, with $\lambda_1$ dedicated to the phase imaging channel and $\lambda_2$ dedicated to the amplitude imaging channel. As an empirical parameter, without loss of generality, we selected $\lambda_2 = \lambda_1 \times 1.28$ and $\lambda_1 + \lambda_2 = 2\lambda$ for our numerical diffractive designs. With this wavelength multiplexing strategy in design II, the amplitude and phase imaging FOVs can be combined into a single FOV – as opposed to 2 spatially separated FOVs as employed by design I shown in **Fig. 1a**. Consequently, the output amplitude and phase images, i.e., $|o_{Amp}(\lambda_2)|$ and $O_{Phase}(\lambda_1)$, can be recorded by the same group of sensor pixels.

As illustrated in **Fig. 1c**, we also developed an additional complex field imager design, referred to as design III, which integrates both space and wavelength multiplexing strategies in constructing the amplitude and phase imaging channels. Specifically, design III incorporates two FOVs that are spatially separated at the output plane (similar to design I) for amplitude and phase imaging, also utilizing two different wavelength channels (akin to design II) to encode the output amplitude/phase images separately.

Following these design configurations (I, II and III) depicted above, we performed their numerical modeling and conducted the training of our diffractive imager models. For this training, we constructed an image dataset comprising 55,000 images of EMNIST handwritten English capital letters, and within each training epoch, we randomly grouped these images in pairs – one representing the amplitude image and another representing the phase image – thereby forming 27,500 training input complex fields. The phase contrast parameter $\alpha_{\text{tr}}$ used for constructing these training input complex fields was set as 1. We utilized deep learning-based optimization with stochastic gradient descent to optimize the thickness values of the diffractive features on the diffractive layers. This training was targeted at minimizing a custom-designed loss function defined by the mean squared error (MSE) between the diffractive imager output amplitude and phase images with respect to their corresponding ground truth. More information about the structural parameters of the diffractive complex field imagers, the specific loss functions employed, and additional aspects of the training methodology can be found in the Methods section.

**Numerical results and quantitative performance analysis of diffractive complex field imagers**

After the training phase, the resulting diffractive layers of our complex field imager models following designs I, II and III are visualized in **Supplementary Figs. S1a**, **S2a** and **Fig. 2a**, respectively, showing their thickness value distributions. To evaluate and quantitatively compare the complex field imaging performances of these diffractive processors, we first conducted blind testing by selecting 10,000 test images from the EMNIST handwritten letter dataset that were never used in the training set and randomly grouped them in pairs to synthesize 5,000 complex test objects. To compare the structural fidelity of the resulting output amplitude and phase images (i.e., $|o_{\text{Amp}}(\lambda)|$ and $O_{\text{Phase}}$) produced by our diffractive complex field imager models, we quantified the peak signal-to-noise ratio (PSNR) metrics between these diffractive output images and their corresponding ground truth (i.e., $A$ and $\phi$). Our results revealed that, for the diffractive imager model using design I that performs space-multiplexed complex field imaging, the amplitude and phase imaging channels provided PSNR values of 16.47±0.96 and 14.90±1.60, respectively, demonstrating a decent imaging performance. Additionally, for the diffractive imager models using designs II (and III), these performance metrics became 16.46±1.02 and 14.98±1.51 (17.04±1.06 and 15.06±1.63), respectively. Therefore, design III demonstrated a notable performance advantage over the other two models in both phase and amplitude imaging channels when both the space and wavelength multiplexing strategies were used. Apart from these quantitative results, we also presented exemplary diffractive output images for the three models of designs I, II and III in **Supplementary Figs. S1b**, **S2b** and **Fig. 2b**, respectively. These visualization results clearly show that our diffractive output images in both amplitude and phase channels present structural similarity to their input ground truth, even though these input complex fields were never seen by our diffractive models before. These analyses demonstrate the *internal generalization* of our diffractive complex field imagers, indicating their capability to process new complex fields that have similar statistical distributions to the training dataset.

We also conducted blind testing of these diffractive complex field imager designs by synthesizing input fields from other datasets where the complex images exhibit distinctly different morphological features compared to the training complex field images. For this purpose, we selected the MNIST handwritten digits[50] and the QuickDraw image[51] datasets, and for each dataset we synthesized 5,000 input complex fields to test our diffractive models blindly. When using the MNIST-based complex field images, the amplitude and phase PSNR values of our diffractive complex field imager models using designs I, II and III were quantified as (16.59±0.71,

15.42±1.28), (16.40±0.68, 15.53±1.25) and (17.05±0.78, 15.59±1.32), respectively. The corresponding diffractive output images for these results are also exemplified in **Supplementary Figs. S1c**, **S2c** and **Fig. 2c**. When testing using input complex fields synthesized from the QuickDraw images, these PSNR values revealed (14.42±0.94, 13.34±1.10), (14.17±1.01, 13.54±1.61) and (14.72±0.97, 13.46±1.13), with exemplary diffractive output images visualized in **Supplementary Figs. S1d**, **S2d** and **Fig. 2d**, respectively. Once again, these PSNR values, along with the visualization results of the output patterns, demonstrate that all our diffractive models (following designs I, II and III) achieved successful reconstructions of the amplitude and phase channel information of the input complex fields, wherein the design III model presented slightly improved performance over the other two designs. Importantly, these analyses demonstrate the *external generalization* capabilities of our diffractive imagers, positioning them as general-purpose complex field imagers that can handle input complex field distributions markedly distinct from those encountered during their training stage.

Next, we quantified the complex field imaging performance of our diffractive models as a function of the phase contrast and spatial resolution of the incoming complex fields. For this analysis, we selected various grating patterns to form our test images, which have different linewidths and are oriented in either horizontal or vertical directions. We first considered using these grating patterns encoded within either the phase or the amplitude channels of the input complex fields, forming phase-only or amplitude-only grating test objects. To be more specific, the phase-only input fields were set to have a uniform distribution within their amplitude channel, while the amplitude-only input fields were set to have their phase channel values set as zero/constant. For both kinds of gratings, we selected their linewidths as $1.5\lambda_m$ or $3\lambda_m$ to generate the grating patterns, and tested the spatial resolution for the amplitude and phase imaging channels using our diffractive models; here $\lambda_m = \lambda$ for the design I model and $\lambda_m = (\lambda_1 + \lambda_2) / 2$ for the designs II and III. For phase-only gratings with linewidths of $3\lambda_m$, we also used different phase contrast parameters $\alpha_{test} \in \{0.25, 0.5, 1\}$ to form grating patterns with different phase contrast so that we can evaluate the sensitivity of phase imaging by our diffractive complex field processors. To better quantify the performance of our diffractive complex field imagers for these test grating patterns, we used a grating image contrast (Q) as our evaluation metric, defined as:

$$Q = \frac{I_{max} - I_{min}}{I_{max} + I_{min}} \qquad (1).$$

The results of using these amplitude- or phase-only grating patterns as input fields to our diffractive models using designs I, II and III are provided in **Supplementary Figs. S3a-b**, **S4a-b** and **Fig. 3a-b**, respectively. Through visual inspection and quantification of grating image contrast Q values, our diffractive imager models were found to resolve most of the amplitude-only grating objects of different linewidths and orientations, with quantified Q values consistently above 0.17. The only exception is that the diffractive model using design II fell short in resolving the horizontal grating patterns with $1.5\lambda_m$ linewidth, achieving Q < 0.1. For the phase-only grating inputs, all three diffractive models succeeded in resolving the gratings with $\alpha_{test} \in \{0.5, 1\}$ and linewidths of $3\lambda_m$, presenting Q values consistently over 0.19. However, when using the phase-only grating inputs with $\alpha_{test} = 0.25$ and linewidths of $3\lambda_m$ or those with $\alpha_{test} = 1$ and linewidths of $1.5\lambda_m$, all of our diffractive models struggle to provide consistently clear grating images, exhibiting relatively poor Q values of ≤ 0.1. These findings reveal that our diffractive imager models exhibit similar performance in imaging resolution and phase sensitivity, providing an amplitude imaging resolution of $>1.5\lambda_m$ for amplitude-only objects and a phase imaging resolution of $\geq 3\lambda_m$ for phase-only objects with $\alpha_{\text{test}} \geq 0.5$. We also calculated the average Q values for different diffractive models using these amplitude- or phase-only grating inputs; the design III model emerges as the most competitive one, presenting average Q values of 0.418 and 0.181 for the amplitude and phase channels, respectively. The suboptimal performance of the design II model, we believe, can primarily be attributed to its utilization of the same output FOV for both the phase and amplitude image formation. This strategy results in the overlap of the diffractive features to serve the two imaging channels, thereby not fully utilizing the degrees of freedom provided by the diffractive layers. This is also corroborated by the visualization of the diffractive layer designs shown in **Supplementary Fig. S2a**: compared to designs I and III, the areas with significant modulation patterns in the design II layers are significantly smaller and more concentrated in the central region, indicating a less efficient utilization of the diffractive degrees of freedom available for optimization, consequently limiting its imaging performance.

In addition to the analyses of spatial resolution and phase sensitivity, we also utilized amplitude- and phase-only grating images to investigate the crosstalk between the amplitude and phase imaging channels of our diffractive complex field imagers. Since the amplitude-only grating inputs have constant/zero phase distributions, the ground truth of their corresponding diffractive output images in the phase channel should have zero intensities, where the residual represents the crosstalk coming from the amplitude channel. Similarly, for the phase-only grating inputs that have a uniform amplitude distribution ($A_{DC}$), their diffractive output images in the amplitude channel should reveal no intensity distributions, with the residual representing the crosstalk coming from the phase channel. As shown by the diffractive output images in **Supplementary Figs. S3a-b**, **S4a-b** and **Fig. 3a-b**, we observe some crosstalk components in the amplitude and phase channel imaging results. To provide a quantitative evaluation of this crosstalk, we used the signal-to-crosstalk ratio (SCR) metric, defined as:

$$SCR_{\text{Phase}} = \frac{\sum O_{\text{Phase} \rightarrow \text{Phase}}}{\sum O_{\text{Amp} \rightarrow \text{Phase}}} \quad (2),$$

$$SCR_{\text{Amp}} = \frac{\sum |o_{\text{Amp} \rightarrow \text{Amp}}|^2}{\sum |o_{\text{Phase} \rightarrow \text{Amp}}|^2} \quad (3),$$

where $O_{\text{Phase} \rightarrow \text{Phase}}$ and $O_{\text{Amp} \rightarrow \text{Phase}}$ denote the resulting output phase image when encoding the same grating pattern within the phase and amplitude channels of the input complex field, respectively; the first term represents the true signal and the latter represents the crosstalk term in Eq. (2). Similarly, $|o_{\text{Amp} \rightarrow \text{Amp}}|$ and $|o_{\text{Phase} \rightarrow \text{Amp}}|$ denote the resulting output amplitude image when encoding the same grating pattern within the amplitude and phase channels of the input complex field, respectively. $\Sigma$ denotes the intensity summation operation across all the pixels. Following these definitions, we quantified the $SCR_{\text{Phase}}$ and $SCR_{\text{Amp}}$ values for all the grating imaging outputs in **Supplementary Figs. S3a-b**, **S4a-b** and **Fig. 3a-b**. These SCR analyses reveal that, for all the diffractive imager models, the grating inputs with $1.5\lambda_m$ linewidth and $\alpha_{\text{test}} = 1$ present a ~30% lower $SCR_{\text{Amp}}$ and a ~53% lower $SCR_{\text{Phase}}$ when compared to their counterparts with $3\lambda_m$ linewidth, revealing that imaging of finer, higher-resolution patterns is more susceptible to crosstalk. Furthermore, we found that an increase in the input phase contrast ($\alpha_{\text{test}}$) leads to more crosstalk in the output amplitude channel, which results in a lower $SCR_{\text{Amp}}$ value; for example, from >3.5 for $\alpha_{\text{test}} = 0.25$ down to 2.5-3 for $\alpha_{\text{test}} = 1$. Additionally, we calculated the average $SCR_{\text{Phase}}$ and $SCR_{\text{Amp}}$ values across these grating images for different diffractive imager models; for the diffractive models using designs I, II and III, the average $SCR_{\text{Amp}}$ values are 2.805, 3.178 and 3.155, respectively, and the average $SCR_{\text{Phase}}$ values are 2.331, 2.262 and 2.252, respectively.

These analyses were performed based on amplitude and phase-only grating objects. Beyond that, we also used complex-valued gratings to further inspect the imaging performance of our diffractive models. Specifically, we created complex test fields that have the same grating patterns encoded in both the amplitude and phase channels. The results reported in the top row of the **Supplementary Figs. S3c**, **S4c** and **Fig. 3c** revealed that all our diffractive imager models are capable of distinctly resolving complex gratings with $3\lambda_{\text{m}}$ linewidth, while being largely able to resolve those with $1.5\lambda_{\text{m}}$ linewidth, albeit with occasional failure. We further created complex fields by orthogonally placing horizontal and vertical gratings, with one of these gratings encoded in the phase channel of the input field and the other encoded in the amplitude channel. As evidenced by the bottom row of **Supplementary Figs. S3c**, **S4c** and **Fig. 3c**, our diffractive models could successfully reconstruct the amplitude and phase patterns of the input complex fields with a grating linewidth of $3\lambda_{\text{m}}$.

**Output power efficiency of diffractive complex field imagers**

To quantify the output diffraction efficiencies of our complex field imagers, we utilized 5,000 test complex fields created from the EMNIST image dataset, and calculated the average diffraction efficiencies of our diffractive complex field imager models. By integrating an additional loss term into our training loss function to balance the complex field imaging performance along with the output diffraction efficiency, we demonstrated the feasibility of increased power efficiency for all three designs (I, II and III), with minimal compromise in the output image quality. The added loss term, denoted as $\mathcal{L}_{\text{Eff}}$, is specifically designed to control and improve the output diffraction power efficiency, with its definition given by:

$$\mathcal{L}_{\text{Eff}} = \mathcal{L}_{\text{Eff,Phase}} + \mathcal{L}_{\text{Eff,Amp}} \tag{4}$$

$$\mathcal{L}_{\text{Eff,Phase}} = \begin{cases} \eta_{\text{th}} - \eta_{\text{Phase}}, & \text{if } \eta_{\text{Phase}} < \eta_{\text{th}} \\ 0, & \text{if } \eta_{\text{Phase}} \geq \eta_{\text{th}} \end{cases} \tag{5}$$

$$\mathcal{L}_{\text{Eff,Amp}} = \begin{cases} \eta_{\text{th}} - \eta_{\text{Amp}}, & \text{if } \eta_{\text{Amp}} < \eta_{\text{th}} \\ 0, & \text{if } \eta_{\text{Amp}} \geq \eta_{\text{th}} \end{cases} \tag{6}$$

where $\eta_{\text{Phase}}$ and $\eta_{\text{Amp}}$ denote the output diffraction power efficiency within $\text{FOV}_{\text{Phase}}$ and $\text{FOV}_{\text{Amp}}$, respectively, with their detailed definition provided in the Methods section. $\eta_{\text{th}}$ refers to the target diffraction efficiency threshold for $\eta$. By minimizing the loss function that incorporates the $\mathcal{L}_{\text{Eff}}$ term, we trained 6 diffractive imager models for each design (I, II and III). For each of these models, we set $\eta_{\text{th}}$ at distinct levels: 0.1%, 0.2%, 0.4%, 0.8%, 1.6% and 3.2%, and trained the respective model to satisfy the specified $\eta_{\text{th}}$. Note that all these new diffractive complex field imager models maintain the same physical architecture as the designs illustrated in **Fig. 1**, and they were trained using the same EMNIST-based complex image dataset. A performance comparison for these models is provided in **Fig. 4**, where their amplitude and phase average PSNR values were calculated across the test set and shown as a function of their average diffraction efficiency values. Taking the architecture of design III as an example, one of our complex field imager designs achieved an output power efficiency of ~0.2% in both amplitude and phase channels, resulting in average PSNR values of 14.72±1.47 and 16.64±1.03 for the two corresponding channels, respectively. An additional model, optimized with a heightened emphasis on the output power efficiency, demonstrated the capability of performing complex field imaging with >0.8% diffraction efficiency in both the phase and amplitude channels, while achieving average amplitude and phase PSNR values of 13.51±1.32 and 16.74±1.05, respectively. A similar trend was also observed for the other models using designs I and II, where a significant increase in the output diffraction efficiency could be achieved with a modest trade-off in the output image quality. Moreover, a comparative assessment of the three different designs under various output diffraction efficiencies reaffirms the overall performance advantage of design III: it presents remarkable advantages over design I in phase imaging while outperforming design II in amplitude imaging. Overall, **Figure 4** serves as a "*designer rule plot*", which offers guidance in selecting suitable diffractive complex field imager models by balancing the phase/amplitude imaging fidelity with output power efficiency according to specific application requirements.

**Experimental validation of diffractive complex field imagers**

We performed experimental validation of our diffractive complex field imagers using the terahertz part of the spectrum, specifically employing the design II configuration as illustrated in **Fig. 1b**; we used $\lambda_1 = 0.75$ mm and $\lambda_2 = 0.8$ mm for the phase and amplitude imaging channels, respectively. We used three diffractive layers for our experimental design, each layer containing 120×120 learnable diffractive features with a lateral size of ~$0.516\lambda_m$ (dictated by the resolution of our 3D printer). The axial spacing between any two adjacent layers (including the diffractive layers and the input/output planes) was chosen as ~$25.8\lambda_m$ (20 mm), resulting in a total axial length of ~$103.2\lambda_m$ for the entire design. As a proof of concept, we designed two experimental models that use different input phase contrast parameters, $\alpha_{\text{exp}} = 1$ and 0.5. These experimental models were trained using a dataset composed of phase-only and amplitude-only objects, which feature randomly generated spatial patterns with binary phase values of {0, $\alpha_{\text{exp}}\pi$} or amplitude values of {0, 1}. In these proof of concept experiments, we did not employ input objects with spatial distributions in both the amplitude and phase channels due to the fabrication challenges of such objects; however, the amplitude-only or phase-only objects used here still share a single common input FOV and are processed by the same diffractive imager. Therefore, this experimental demonstration serves as an effective proof of our all-optical complex field imaging framework, which has never been demonstrated before in prior works.

After the training, the resulting layer thickness profiles of the diffractive models with $\alpha_{\text{exp}} = 1$ and 0.5 are visualized in **Fig. 5a** and **d**, respectively. These diffractive layers were fabricated using 3D printing, with their corresponding photographs showcased in **Fig. 5b** and **e**. Additionally, we constructed phase-only or amplitude-only test objects, which were never seen by the trained diffractive models. The phase-only test objects were fabricated by 3D printing layers with spatially varying height profiles representing the phase distributions, and the amplitude-only objects were created by padding aluminum foils onto 3D-printed flat layers to delineate the amplitude patterns. In our proof-of-concept experiments, these objects were designed to have 5×5 pixels, each featuring a size of 4.8 mm (~$6.19\lambda_m$). As shown in **Fig. 6b**, the printed diffractive layers and input complex objects were assembled using a custom 3D-printed holder to ensure that their relative positions follow our numerical design. In our experiments, we employed a THz source operating at $\lambda_1 = 0.75$ mm and $\lambda_2 = 0.8$ mm, and used a detector to measure the intensity distribution at the output plane, yielding the output amplitude and phase images. The photograph and schematic of our experimental setup are provided in **Fig. 6a** and **c**, respectively. Further details related to the experiment are provided in the Methods section.

The experimental results for these two models are shown in **Fig. 5c** and **f**, where the output amplitude and phase images present a good agreement with their numerically simulated counterparts, also aligning well with the input ground truth images. These experimental results demonstrate the feasibility of our 3D fabricated diffractive complex field imager to accurately image the amplitude and phase distributions of the input objects; these results also represent the first demonstration of all-optical complex field imaging achieved through a single diffractive processor.

## Discussion

The numerical analyses and experimental validation presented in our work showcased a compact complex field imager design through deep learning-based optimization of diffractive surfaces. We explored three variants of this design strategy, with comparative analyses indicating that the design employing spatial and wavelength multiplexing (design III) achieves the best balance between the complex field imaging performance and diffraction efficiency, albeit with a minor increase in hardware complexity. Leveraging the all-optical information processing capabilities of multiple spatially engineered diffractive layers, diffractive complex field imagers reconstruct the amplitude and phase distributions of the input complex field in a complete end-to-end manner, without any digital image recovery algorithm, setting it apart from other designs in the existing literature for similar applications. This capability enables direct recording of the amplitude and phase information in a single snapshot using an intensity-only sensor array, which obviates the need for additional computational processing in the back-end, thereby significantly enhancing the frame rate and reducing the latency of the imaging process.

In our previous research, we developed diffractive processor designs tailored for imaging either amplitude distributions of amplitude-only objects[38] or phase distributions of phase-only objects[40,48,49]. However, these designs would become ineffective for imaging complex objects with independent and non-uniform distributions in the amplitude and phase channels. In this work, we have overcome this limitation by training our diffractive imager designs using complex objects with random combinations of amplitude and phase patterns, thus allowing a single imager device to effectively generalize to complex optical fields with various distributions in the amplitude and phase channels.

The diffractive complex field imager designs that we presented also exhibit certain limitations. Our results revealed residual errors in their targeted operations, particularly manifesting as crosstalk coming from the amplitude channel into the phase channel. This suggests that the actual phase-to-intensity transformation represented by our diffractive imager, while effective, is an approximation with errors that are dependent on the object amplitude distribution. The mitigation approach for this limitation might involve further enhancement of the information processing capacity of our diffractive imagers, which can be achieved through employing a larger number of diffractive layers (forming a deeper diffractive architecture), thus increasing the overall number of diffractive features/neurons that are efficiently utilized[52]. Additionally, we believe another performance improvement strategy could be to increase the lateral distance between the two output FOVs dedicated to the phase and amplitude channels, thereby allowing the trainable diffractive features to better specialize for the individual tasks of phase/amplitude imaging; this approach, however, would increase the size of the output FOV of the focal plane array and also demand larger diffractive layers.

Moreover, in our experimental results, we observed the emergence of noise patterns within certain regions, which did not exist in our numerical simulations. This discrepancy can be attributed to potential misalignments and fabrication imperfections in the diffractive layers that are assembled. A mitigation strategy could be to perform "vaccination" of these diffractive imager models, which involves modeling these errors as random variables and incorporating them into the physical forward model during the training process[35,37,53]. This has been proven effective in providing substantial resilience against misalignment errors for diffractive processors, exhibiting a noticeably better match between the numerical and experimental results[35,37,53].

## Methods

**Numerical forward model of a diffractive complex field imager.** In our numerical implementation, the transmissive layers within the diffractive complex field imager were modeled as thin dielectric optical modulation elements with spatially varying thickness profiles. For the $l^{th}$ diffractive layer, the complex-valued transmission coefficient of its $i^{th}$ feature at a spatial location $(x_i, y_i, z_l)$ was defined depending on the illumination wavelength ($\lambda$):

$$t^l(x_i, y_i, z_l, \lambda) = a^l(x_i, y_i, z_l, \lambda) \exp\left(j\phi^l(x_i, y_i, z_l, \lambda)\right) \quad (7),$$

where $a(x_i, y_i, z_l, \lambda)$ and $\phi(x_i, y_i, z_l, \lambda)$ denote the amplitude and phase coefficients, respectively. The free-space propagation of complex fields between diffractive layers was modeled through the Rayleigh-Sommerfeld diffraction equation[32]:

$$w_i^l(x, y, z, \lambda) = \frac{z - z_l}{r^2}\left(\frac{1}{2\pi r} + \frac{1}{j\lambda}\right) \exp\left(\frac{j2\pi r}{\lambda}\right) \quad (8),$$

where $w_i^l(x, y, z, \lambda)$ represents the complex field at the $i^{th}$ diffractive feature of the $l^{th}$ layer at location $(x, y, z)$. $r = \sqrt{(x - x_i)^2 + (y - y_i)^2 + (z - z_l)^2}$ and $j = \sqrt{-1}$. Based on Eq. (8), $w_i^l(x, y, z, \lambda)$ can be viewed as a secondary wave generated from the source at $(x_i, y_i, z_l)$. As a result, the optical field modulated by the $i^{th}$ diffractive feature of the $l^{th}$ layer ($l \geq 1$, treating the input object plane as the $0^{th}$ layer), $u^l(x_i, y_i, z_l, \lambda)$, can be written as:

$$u^l(x_i, y_i, z_l, \lambda) = t^l(x_i, y_i, z_l, \lambda) \cdot \sum_{k \in N} u^{l-1}(x_k, y_k, z_{l-1}, \lambda) \cdot w_i^{l-1}(x_i, y_i, z_l, \lambda) \quad (9),$$

where $N$ denotes the number of diffractive features on the $(l-1)^{th}$ diffractive layer and $z_*$ represents the location of the $*^{th}$ layer in the z direction parallel to the optical axis. The amplitude and phase components of the complex transmittance of the $i^{th}$ feature of diffractive layer $l$, i.e., $a^l(x_i, y_i, z_l, \lambda)$ and $\phi^l(x_i, y_i, z_l, \lambda)$ in Eq. (7), were defined as a function of the material thickness over the region of that diffractive feature, $h_i^l$, as follows:

$$a^l(x_i, y_i, z_l, \lambda) = \exp\left(-\frac{2\pi \kappa_d(\lambda) h_i^l}{\lambda}\right) \quad (10),$$

$$\phi^l(x_i, y_i, z_l, \lambda) = (n_d(\lambda) - n_{air})\frac{2\pi h_i^l}{\lambda} \quad (11),$$

Here the parameters $n_d(\lambda)$ and $\kappa_d(\lambda)$ represent the refractive index and the extinction coefficient of the diffractive layer material, respectively. These parameters correspond to the real and imaginary parts of the complex-valued refractive index, denoted as $\tilde{n}_d(\lambda)$, such that $\tilde{n}_d(\lambda) = n_d(\lambda) + j\kappa_d(\lambda)$. We determined the values of $\tilde{n}_d(\lambda)$ and $\kappa_d(\lambda)$ through experimental characterization of the dispersion properties of the diffractive layer materials, and their values are visualized in **Supplementary Fig. S5**. The trainable thickness values of the diffractive features $h_i^l$ were limited within the range of $[h_{\min}, h_{\max}]$, representing the learnable parameters of our diffractive complex field imagers. For training the diffractive imager models used for numerical analyses, the values of $h_{\min}, h_{\max}$ were selected as 0.2 and 1.2 mm, respectively. For training the diffractive imager models used for experimental validation, the values of $h_{\min}, h_{\max}$ were selected as 0.4 and 1.4 mm, respectively.

**Experimental terahertz set-up.** For our proof-of-concept experiments, we fabricated both the diffractive layers and the test objects using a 3D printer (PR110, CADworks3D). The phase objects were fabricated with spatially varying thickness profiles to define their phase distributions. The amplitude objects were printed to have a uniform thickness and then manually coated with aluminum foil to define the light-blocking areas, while the uncoated sections formed the transmission areas, resulting in the creation of the desired amplitude profiles for test objects. Additionally, we 3D-printed a holder using the same 3D printer, which facilitated the assembly of the printed diffractive layers and input objects to align with their relative positions as specified in our numerical design. To more precisely control the beam profile for the illumination of the complex input objects, we 3D printed a square-shaped aperture of 5×5 mm and padded the area around it with aluminum foil. The pinhole was positioned 120 mm away from the object plane in our experiments. This pinhole serves as an input spatial filter to clean the beam originating from the source.

To test our fabricated diffractive complex field design, we employed a THz continuous-wave scanning system, with its schematic presented in **Fig. 6c**. To generate the incident terahertz wave, we used a WR2.2 modular amplifier/multiplier chain (AMC) followed by a compatible diagonal horn antenna (Virginia Diode Inc.) as the source. Each time, we transmitted a 10 dBm sinusoidal signal at frequencies of 11.111 or 10.417 GHz (fRF1) to the source, which was then multiplied 36 times to generate output radiation at continuous-wave (CW) radiation at frequencies of 0.4 or 0.375 THz, respectively, corresponding to the illumination wavelengths of 0.75 and 0.8 mm used for the phase and amplitude imaging tasks, respectively. The AMC output was also modulated with a 1 kHz square wave for lock-in detection. We positioned the source antenna to be very close to the 3D-printed spatial pinhole filter, such that the illumination power input to the system could be maximized. Next, using a single-pixel detector with an aperture size of ~0.1 mm, we scanned the resulting diffraction patterns at the output plane of the diffractive complex field imager at a step size of 0.8 mm. This detector was mounted on an XY positioning stage constructed from linear motorized stages (Thorlabs NRT100) and aligned perpendicularly for precise control of the detector's position. For illumination at $\lambda_1 = 0.75$ mm or $\lambda_2 = 0.8$ mm, a 10-dBm sinusoidal signal was also generated at 11.083 or 10.389 GHz (fRF2), respectively, as a local oscillator and sent to the detector to down-convert the output signal to 1 GHz. The resulting signal was then channeled into a low-noise amplifier (Mini-Circuits ZRL-1150-LN+) with an 80 dBm gain, followed by a bandpass filter at 1 GHz (± 10 MHz) (KL Electronics 3C40-1000/T10-O/O), effectively mitigating noise from undesired frequency bands. Subsequently, the signal passed through a tunable attenuator (HP 8495B) for linear calibration before being directed to a low-noise power detector (Mini-Circuits ZX47-60). The voltage output from the detector was measured using a lock-in amplifier (Stanford Research SR830), which utilized a 1 kHz square wave as the reference signal. The readings from the lock-in amplifier were then calibrated into a linear scale. In our post-processing, we further applied linear interpolation to each intensity field measurement to align with the pixel size of the output FOV used in the design phase. This process finally resulted in the output measurement images presented in **Fig. 5c** and **f**.

31. Li, L. *et al.* Single-Shot Wavefront Sensing with Nonlocal Thin Film Optical Filters. *Laser Photonics Rev.* **17**, 2300426 (2023).

32. Lin, X. *et al.* All-optical machine learning using diffractive deep neural networks. *Science* **361**, 1004–1008 (2018).

33. Mengu, D., Luo, Y., Rivenson, Y. & Ozcan, A. Analysis of Diffractive Optical Neural Networks and Their Integration With Electronic Neural Networks. *IEEE J. Sel. Top. Quantum Electron.* **26**, 1–14 (2020).

34. Veli, M. *et al.* Terahertz pulse shaping using diffractive surfaces. *Nat. Commun.* **12**, 37 (2021).

35. Li, J. *et al.* Spectrally encoded single-pixel machine vision using diffractive networks. *Sci. Adv.* **7**, eabd7690 (2021).

36. Işıl, Ç. *et al.* Super-resolution image display using diffractive decoders. *Sci. Adv.* **8**, eadd3433 (2022).

37. Bai, B. *et al.* To image, or not to image: class-specific diffractive cameras with all-optical erasure of undesired objects. *eLight* **2**, 14 (2022).

38. Li, J. *et al.* Unidirectional imaging using deep learning–designed materials. *Sci. Adv.* **9**, eadg1505 (2023).

39. Li, Y. *et al.* Universal Polarization Transformations: Spatial Programming of Polarization Scattering Matrices Using a Deep Learning-Designed Diffractive Polarization Transformer. *Adv. Mater.* **35**, 2303395 (2023).

40. Shen, C.-Y., Li, J., Mengu, D. & Ozcan, A. Multispectral Quantitative Phase Imaging Using a Diffractive Optical Network. *Adv. Intell. Syst.* 2300300 (2023) doi:10.1002/aisy.202300300.
16

**Supplementary Materials:**
- Supplementary Figures S1-S5
- Training loss functions and quantification metrics
- Implementation details of diffractive complex field imagers

# Figures

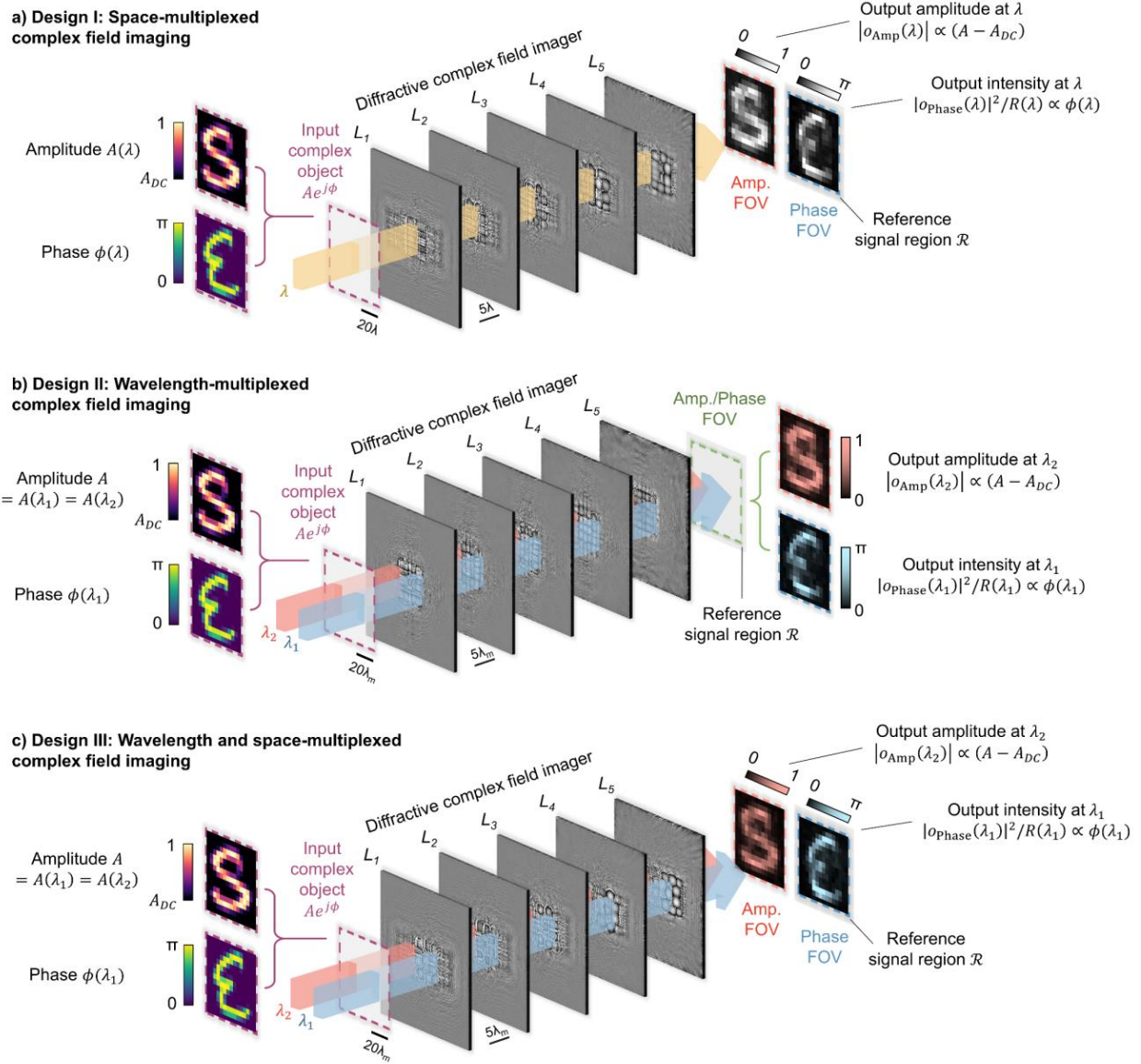

**Figure 1. Schematics for different designs of our diffractive complex field imager. a,** Illustration showing a spatially multiplexed design for the diffractive complex field imager (design I), which performs imaging of the amplitude and phase distributions of the input complex object simultaneously by channeling the output amplitude and phase images onto two spatially separate FOVs at the output plane, i.e., the amplitude and phase FOVs (or $\text{FOV}_{\text{Amp}}$ and $\text{FOV}_{\text{Phase}}$). **b,** Illustration for an alternative design of the diffractive complex field imager using wavelength multiplexing (design II), wherein the output amplitude and phase profiles are directly measured using a common output FOV but at different wavelengths, i.e., $\lambda_1$ and $\lambda_2$, respectively. **c,** Illustration for design III of the diffractive complex field imager, wherein the output amplitude and phase images are measured using two spatially separate FOVs ($\text{FOV}_{\text{Amp}}$ and $\text{FOV}_{\text{Phase}}$) and also at different wavelengths ($\lambda_1$ and $\lambda_2$, respectively).

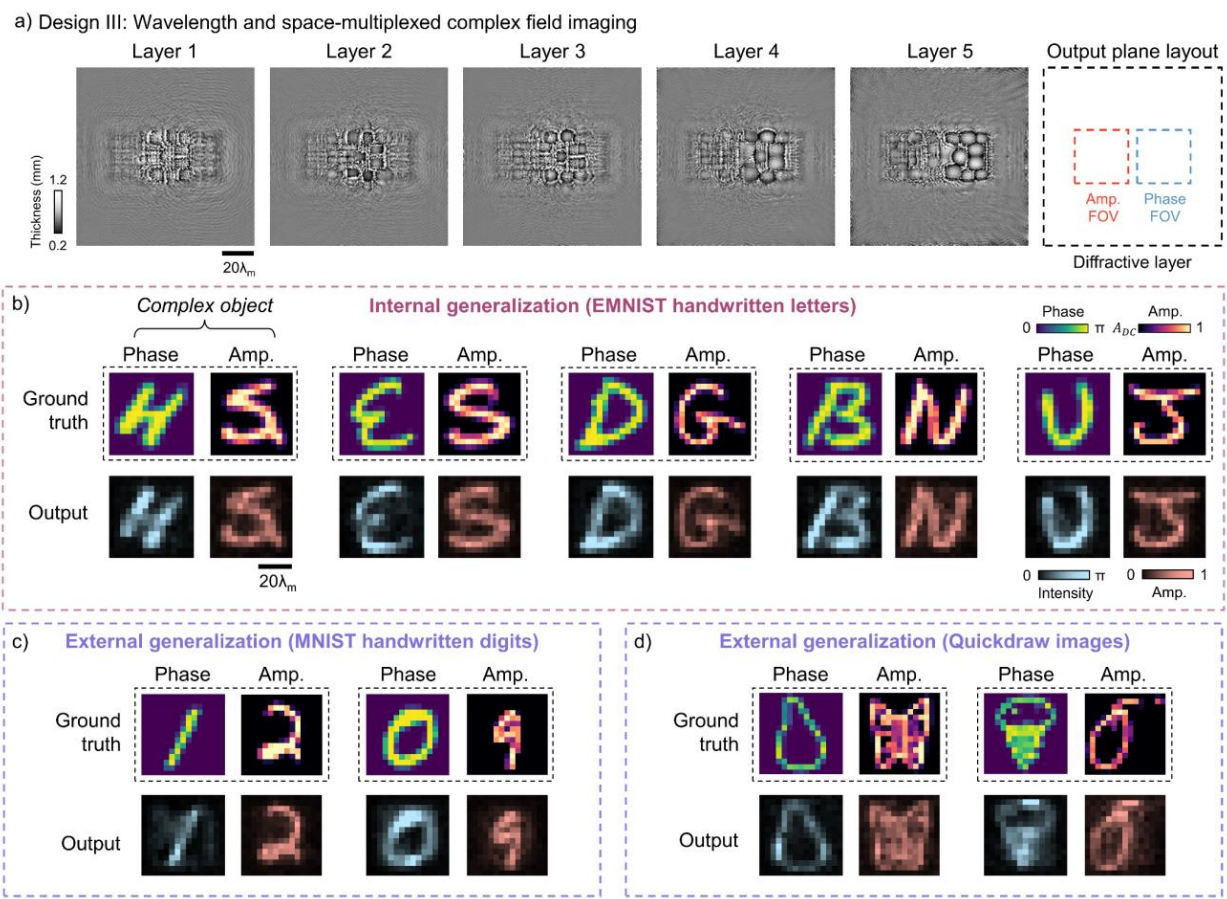

**Figure 2. Blind testing results of the diffractive complex field imager using design III. a,** Thickness profile of the trained layers of the diffractive complex field imager following the design III in Fig. 1c. The layout of the amplitude and phase FOVs in comparison to the size of a diffractive layer is also provided. **b,** Exemplary blind testing input complex objects never seen by the diffractive imager model during its training, along with their corresponding output amplitude and phase images. **c and d,** Same as (b), except that the testing images are taken from the MNIST and QuickDraw datasets, respectively, demonstrating external generalization to image datasets with different structural distributions.

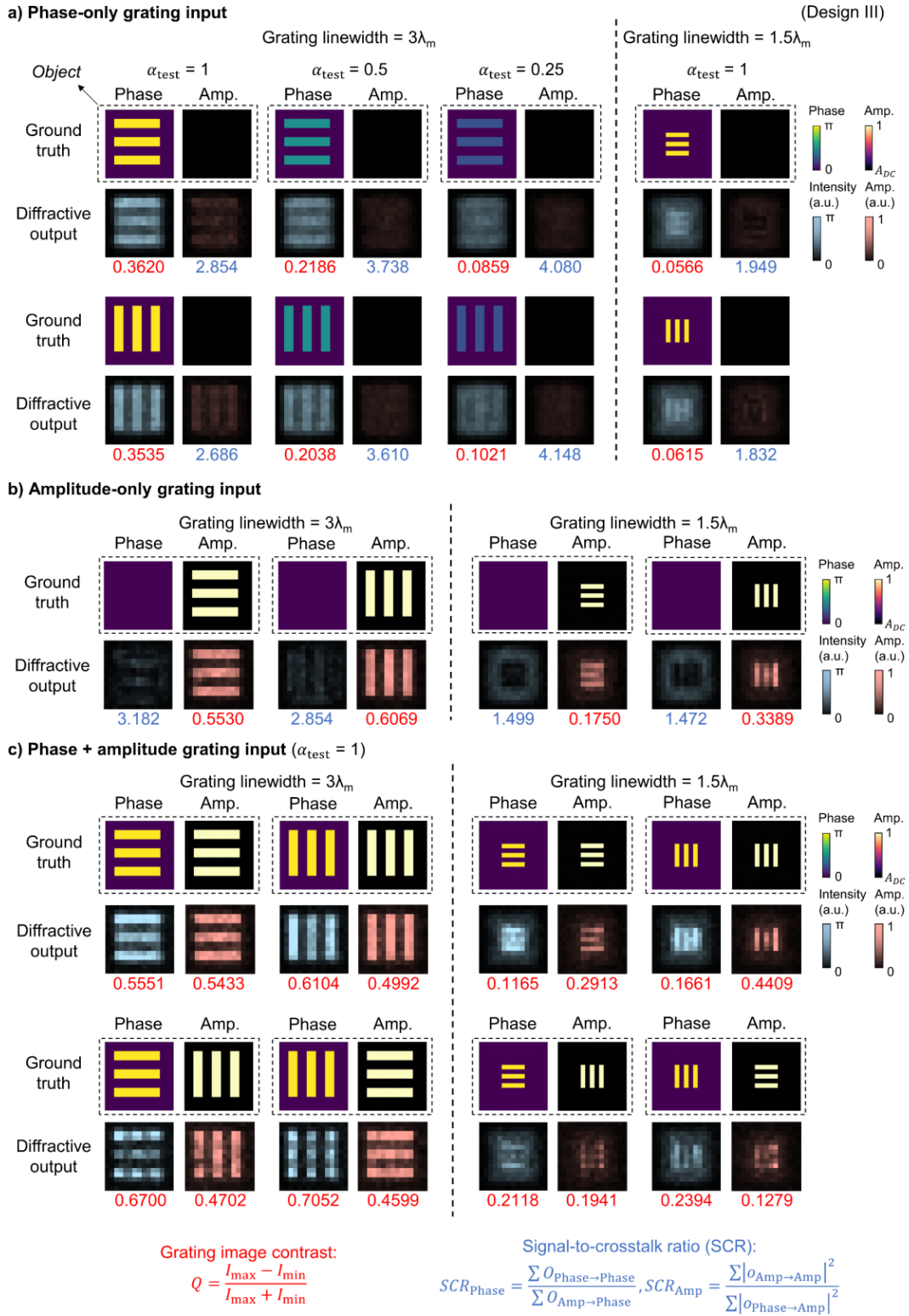

**Figure 3. Performance analysis of the diffractive complex field imager model shown in Fig. 2. a,** Imaging results using phase-only gratings as input fields. The binary phase grating patterns encoded within the phase channel of the input objects are shown and compared with the resulting

output amplitude and phase images produced by our diffractive imager (i.e., $|o_{\text{Amp}}(\lambda)|$ and $O_{\text{Phase}}$). For each diffractive output image, the grating image contrast Q and SCR values were quantified and shown in red and blue numbers, respectively. **b,** Same as in (a), except that the amplitude-only gratings are used as input fields. **c**, Imaging results using complex grating objects as input fields. These grating test objects include ones with the same grating patterns encoded in both the amplitude and phase channels (top), as well as ones where horizontal and vertical gratings are orthogonally placed, with one encoded in the phase channel of the input field and the other encoded in the amplitude channel (bottom).

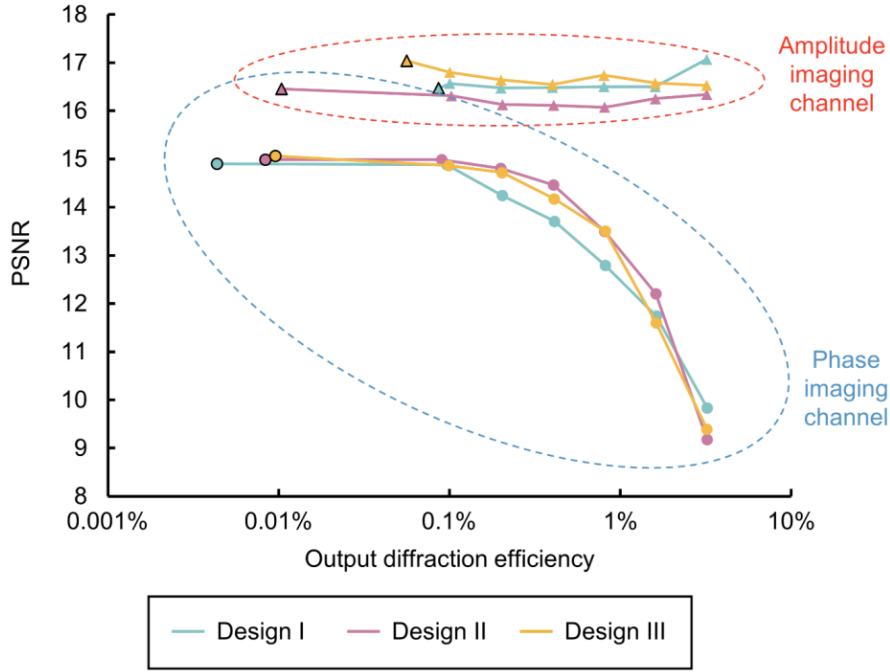

**Figure 4. The trade-off between the complex field imaging performance and the output diffraction efficiency of diffractive complex field imagers.** The PSNR on the y-axis reflects the mean value computed over the entire 5,000 complex test objects derived from the EMNIST dataset. The data points with black borders correspond to the diffractive imager models trained exclusively using the structural fidelity loss function while disregarding diffraction efficiency, i.e., the ones shown in **Supplementary Figs. S1a**, **S2a** and **Fig. 2a**. The other data points originate from models trained with a diffraction efficiency-related penalty term, as defined in Eq. (12). These models were trained using varying target diffraction efficiency thresholds ($\eta_{th}$), specifically set at 0.1%, 0.2%, 0.4%, 0.8%, 1.6% and 3.2% corresponding to the data points from left to right on the plot, demonstrating the trade-off between the imaging performance and the output diffraction efficiency.

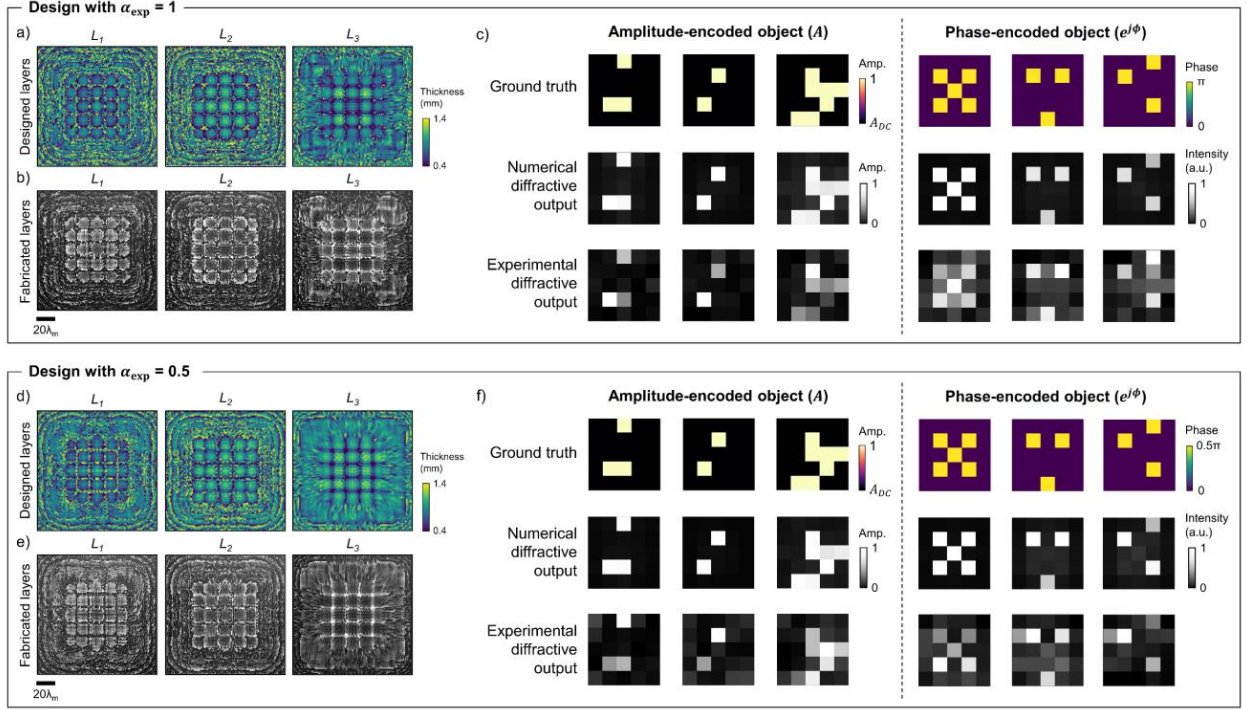

**Figure 5. Experimental results for 3D-printed diffractive complex field imagers. a,** The learned thickness profiles of the layers ($L_1$, $L_2$ and $L_3$) of the diffractive imager model trained with $\alpha_{\text{exp}} = 1$. **b,** Photographs of the 3D-printed diffractive layers in (a). **c,** Experimental results of the diffractive imager shown in (a), compared with their corresponding numerical simulation results and ground truth images. **d,** The learned thickness profiles of the layers ($L_1$, $L_2$ and $L_3$) of the diffractive imager model trained with $\alpha_{\text{exp}} = 0.5$. **e,** Photographs of the 3D-printed diffractive layers in (d). **f,** Experimental results of the diffractive imager shown in (d), compared with their corresponding numerical simulation results and ground truth images.

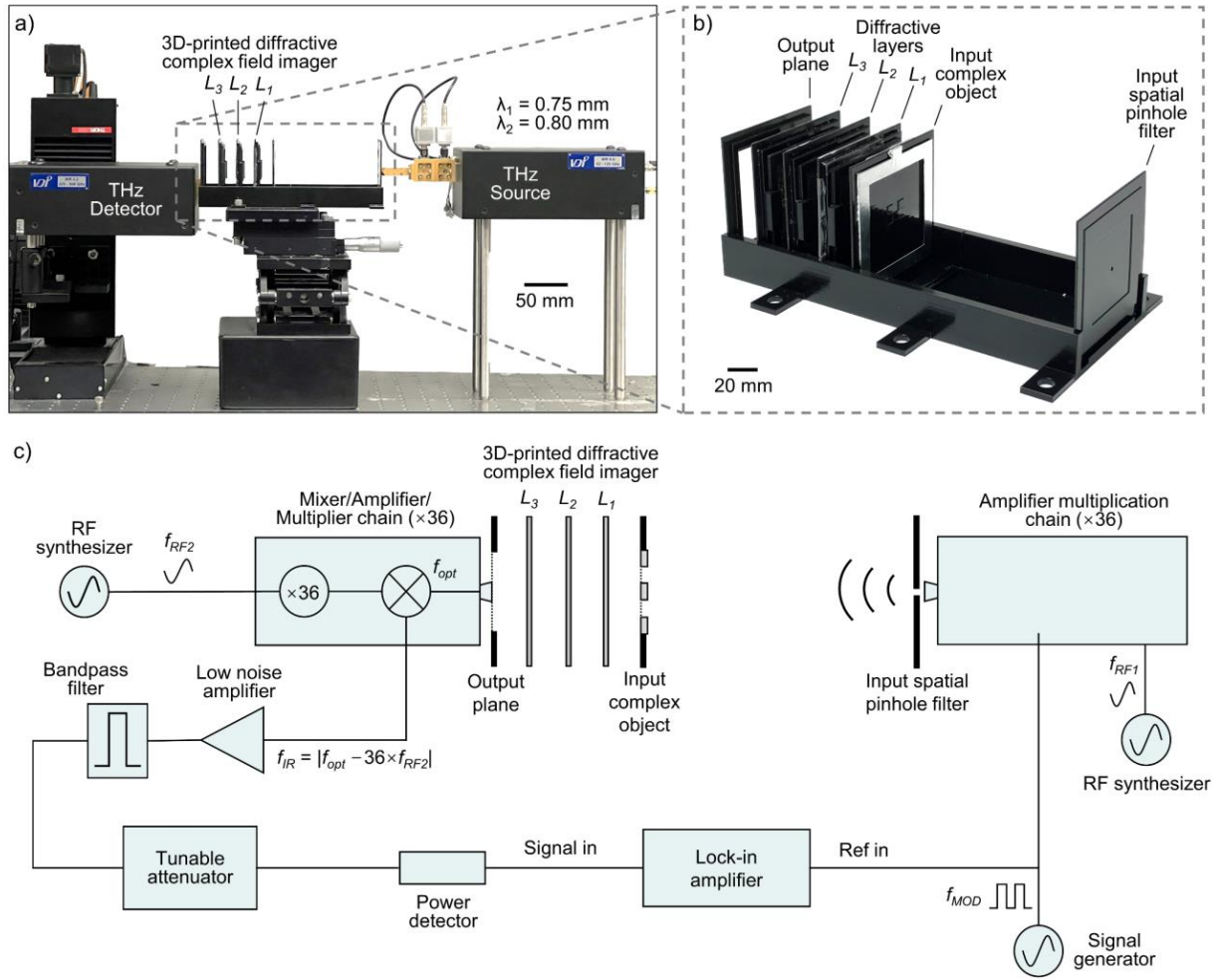

**Figure 6. Experimental set-up for diffractive complex field imagers. a,** Photograph of the experimental set-up, including a 3D-fabricated diffractive complex field imager. **b,** Photographs of the 3D printed diffractive complex field imager. **c,** Schematic diagram of the continuous-wave terahertz imaging set-up.